\pdfoutput=1

\documentclass[aps,prl,10pt,twocolumn,showpacs,superscriptaddress]{revtex4-1}
\usepackage{amsmath}
\usepackage{latexsym}
\usepackage{amssymb}
\usepackage{graphics,epstopdf}

\newcommand{\bra}[1]{\langle #1 |}
\newcommand{\ket}[1]{| #1 \rangle}

\newcommand{\tr}{\mbox{tr}}

\begin{document}

\author{Kavan Modi}
\email{kavmodi@gmail.com}
\affiliation{Centre for Quantum Technologies, National University of Singapore, Singapore}

\author{Tomasz Paterek}
\affiliation{Centre for Quantum Technologies, National University of Singapore, Singapore}

\author{Wonmin Son}
\affiliation{Centre for Quantum Technologies, National University of Singapore, Singapore}

\author{Vlatko Vedral}
\affiliation{Centre for Quantum Technologies, National University of Singapore, Singapore}
\affiliation{Department of Physics, National University of Singapore, Singapore}
\affiliation{Clarendon Laboratory, University of Oxford, Oxford UK}

\author{Mark Williamson}
\affiliation{Centre for Quantum Technologies, National University of Singapore, Singapore}

\title{Unified view of quantum and classical correlations}
\date{\today}

\begin{abstract}
We discuss the problem of separation of total correlations in a given quantum state into entanglement, dissonance, and classical correlations using the concept of relative entropy as a distance measure of correlations. This allows us to put all correlations on an equal footing. Entanglement and dissonance, whose definition is introduced here, jointly belong to what is known as quantum discord. Our methods are completely applicable for multipartite systems of arbitrary dimensions. We investigate additivity relations between different correlations and show that dissonance may be present in pure multipartite states.
\end{abstract}

\maketitle


\emph{Introduction.}---Quantum systems are correlated in ways inaccessible to classical objects. A distinctive quantum feature of correlations is quantum entanglement \citep{PhysRev.47.777,Schrodinger:1935eq,PhysRev.48.696}. Entangled states are nonclassical as they cannot be prepared with the help of local operations and classical communication (LOCC) \cite{horodecki:865}. However, it is not the only aspect of nonclassicality of correlations due to the nature of operations allowed in the framework of LOCC. To illustrate this, one can compare a classical bit with a quantum bit; in the case of full knowledge about a classical bit, it is completely described by one of two locally distinguishable states, and the only allowed operations on the classical bit are to keep its value or flip it. 
To the contrary, quantum operations can prepare quantum states that are indistinguishable for a given measurement.
Such operations and classical communication can lead to separable states (those which can be prepared via LOCC) which are mixtures of locally indistinguishable states. These states are nonclassical in the sense that they cannot be prepared using classical operations on classical bits.

Recent measures of nonclassical correlations are motivated by different notions of classicality and operational means to quantify nonclassicality \cite{henderson01a, PhysRevLett.88.017901, PhysRevLett.89.180402, PhysRevA.72.032317, luo:022301}. Quantum discord has received much attention in studies involving thermodynamics and correlations \cite{zurek, PhysRevA.71.062307, DillenschneiderLutz}, positivity of dynamics \cite{Rodriguez07a, shabanilidar09a}, quantum computation \cite{dattashaji, Lanyon, datta, cui}, broadcasting of quantum states \cite{piani, pianib}, dynamics of discord \cite{werlangPRA, fanchini, jhcole}, and volume of discord \cite{modi, ferraro}. Most of these works are limited to studies of bipartite correlations only as the concept of discord, which relies on the definition of mutual information, is not defined for multipartite systems. In some of the studies, it is also desirable to compare various notions of quantum correlations. It is well known that the different measures of quantum correlation are not identical and conceptually different. For example, the discord does not coincide with entanglement and a direct comparison of two notions is rather meaningless. Therefore, an unified classification of correlations is in demand.

\begin{figure}[!ht]
\resizebox{6 cm}{5.265 cm}{\includegraphics{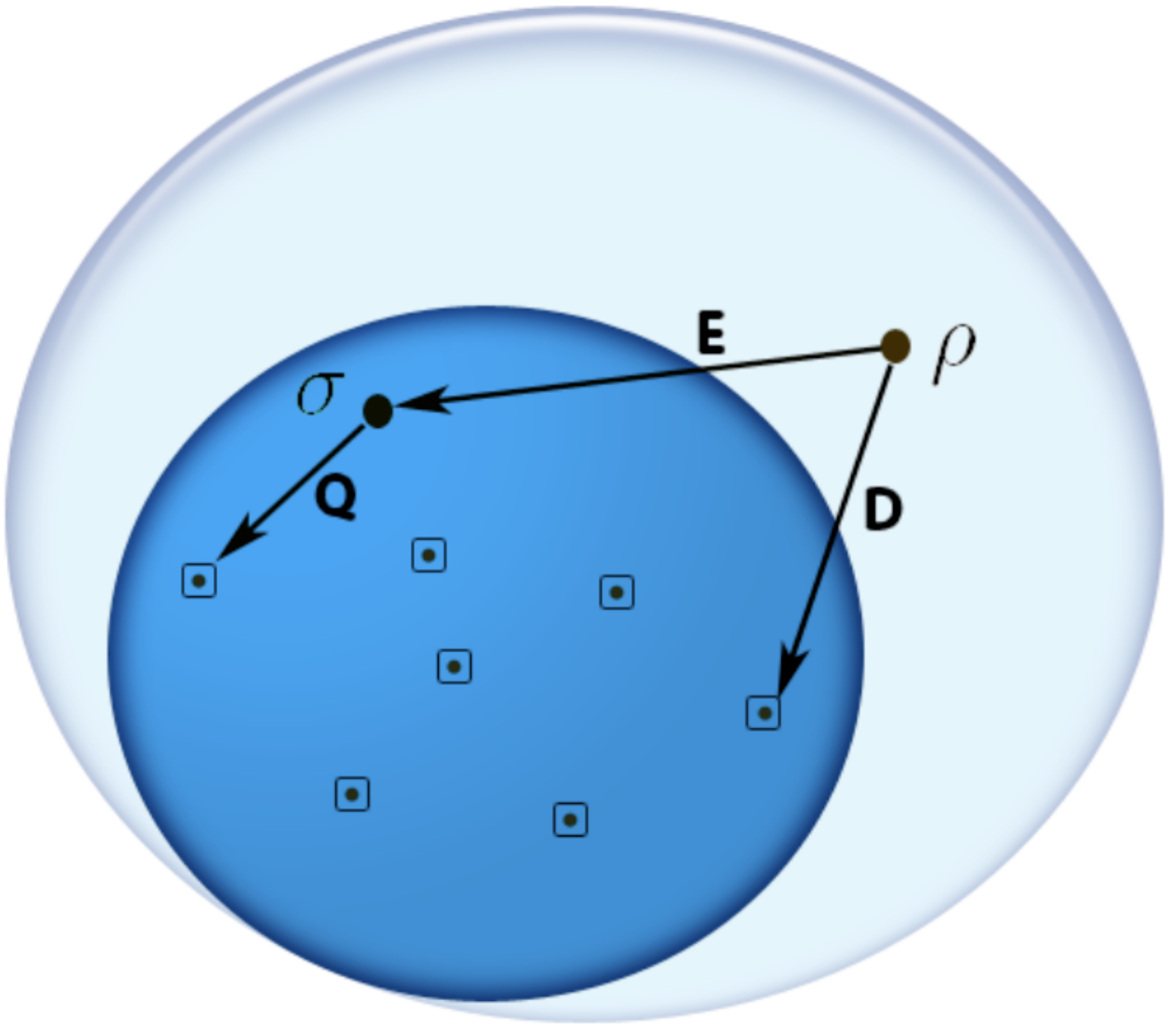}}
\caption{\emph{Correlations as a distance.}  The large ellipse represents the set of all states with the set of separable states in the smaller ellipse. The squares represent the set of classical states, and the dots within the squares are the sets of product states. $\rho$ is an entangled state and $\sigma$ is the closest separable state. The correlations are entanglement, $E$, discord, $D$, and dissonance, $Q$.\label{CORRELATIONS}}
\end{figure}

In this Letter, we resolve these two issues by introducing a measure for classical and nonclassical correlations for quantum states which is applicable for multipartite systems. Our measure of correlations is based on the idea that a distance from a given state to the closest state without the desired property (e.g. entanglement or discord) is a measure of that property. For example, the distance to the closest separable state is a meaningful measure of entanglement. If the distance is measured with relative entropy, the resulting measure of entanglement is the relative entropy of entanglement \cite{VPRK,VP}. In this Letter, using relative entropy we define measures of nonclassical correlations as a distance to the closest classical states, though many other distance measures can serve just as well. Since all the distances are measured with relative entropy, this provides a consistent way to compare different correlations, such as entanglement, discord, classical correlations, and \emph{quantum dissonance}, a new quantum correlation that may be present in separable states. Dissonance is a similar notion to discord, but it excludes entanglement. 
We give formulae for various correlations and show additivity and subadditivity of correlations. We find that a pure multipartite state, the $W$ state, contains dissonance along with entanglement unlike the general bipartite pure state case.  Finally, we compare our results with the \emph{original definition of discord} \cite{henderson01a, PhysRevLett.88.017901} and \emph{measurement induced disturbance} \cite{luo:022301}.

\emph{Definitions.}---We begin by providing the definitions of the states discussed in this Letter. A product state of $N$-partite system, a state with no correlations of any kind, has the form of
$\pi=\pi_1 \otimes \dots \otimes \pi_N$,
where $\pi_n$ is the reduced state of the $n$th subsystem. The set of product states, $\mathcal{P}$, is not a convex set in the sense a mixture of product states may not be another product state. The set of classical states, $\mathcal{C}$, contains mixtures of locally distinguishable states
$\chi = \sum_{k_n} p_{k_1 \dots k_N} \ket{k_1\dots k_N}\bra{k_1 \dots k_N}
= \sum_{\vec{k}} p_{\vec{k}} \ket{\vec{k}}\bra{\vec{k}}$,
where $p_{\vec k} $ is a joint probability distribution and local states $\ket{k_n}$ span an orthonormal basis. The correlations of these states are identified as classical correlations \cite{henderson01a, PhysRevLett.88.017901, PhysRevLett.89.180402,li:024303}. Note that $\mathcal{C}$ is not a convex set; mixing two classical states written in different bases can give rise to a nonclassical state. The set of separable states, $\mathcal{S}$, is convex and contains mixtures of the form
$\sigma = \sum_{i} p_i \pi_1^{(i)} \otimes \dots \otimes \pi_N^{(i)}$.
These states can be prepared using only local quantum operations and classical communication \cite{PhysRevA.40.4277} and can possess nonclassical features \cite{henderson01a, PhysRevLett.88.017901}. The set of product states is a subset of the set of classical states which in turn is a subset of the set of separable states.
Finally, entangled states are all those which do not belong to the set of separable states.  The set of entangled states, $\mathcal{E}$, is not a convex set either.

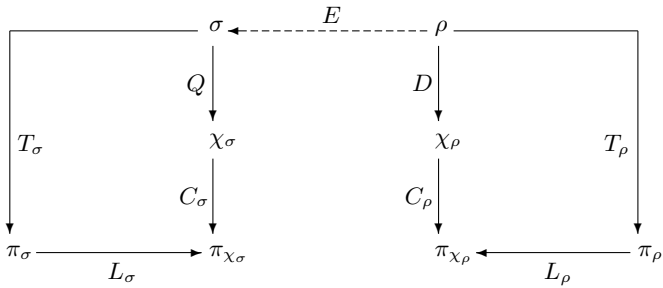
\begin{figure}[t]
	\setlength{\unitlength}{5cm}
	\begin{picture}(2.2,0.75)
	\put(0.55,0.7){$\sigma$} 
	\put(0.52,.7){\line(-1,0){.50}} 
	\put(0.02,.7){\vector(0,-1){.54}} 
	\put(0.04,0.38){$T_\sigma$}
	\put(0.56,.66){\vector(0,-1){.2}} 
	\put(.49,0.54){$Q$} 
	\put(1.15,0.7){$\rho$} 
	\multiput(1.11,.7)(-.03,0){16}{\line(1,0){.02}}
	\put(.65,.7){\vector(-1,0){.05}} 
	\put(.85,0.72){$E$} 
	\put(1.16,.66){\vector(0,-1){.2}} 
	\put(1.09,0.54){$D$} 
	\put(1.2,0.7){\line(1,0){.49}} 
	\put(1.69,.7){\vector(0,-1){.54}} 
	\put(1.6,0.38){$T_\rho$}
	\put(0.55,0.4){$\chi_\sigma$} 
	\put(0.56,.36){\vector(0,-1){.2}} 
	\put(0.47,0.24){$C_\sigma$} 
	\put(1.15,0.4){$\chi_\rho$} 
	\put(1.16,.36){\vector(0,-1){.2}} 
	\put(1.07,0.24){$C_\rho$} 
	\put(0.01,0.1){$\pi_\sigma$} 
	\put(0.09,0.11){\vector(1,0){.44}} 
	\put(0.28,0.04){$L_\sigma$} 
	\put(0.55,0.1){$\pi_{\chi_\sigma}$} 
	\put(1.15,0.1){$\pi_{\chi_\rho}$} 
	\put(1.69,0.1){$\pi_\rho$} 
	\put(1.67,0.11){\vector(-1,0){.41}} 
	\put(1.44,0.04){$L_\rho$} 
	\end{picture}
\caption{\label{ALLSTATES}\emph{Correlations in a quantum state.} An arrow from $x$ to $y$, $x \to y$, indicates that $y$ is the closest state to $x$ as measured by the relative entropy $S(x||y)$. The state $\rho\in\mathcal{E}$ (the set of entangled states), $\sigma\in\mathcal{S}$ (the set of separable states), $\chi\in\mathcal{C}$ (the set of classical states), and $\pi\in\mathcal{P}$ (the set of product states).  The distances are entanglement, $E$, quantum discord, $D$, quantum dissonance, $Q$, total mutual information, $T_\rho$ and $T_\sigma$, and classical correlations, $C_\sigma$ and $C_\rho$. All relative entropies, except for entanglement, reduce to the differences in entropies of $y$ and $x$, $S(x||y)=S(y)-S(x)$. With the aid of $L_\rho$ and $L_\sigma$ the closed path are additive, i.e. Eq. \ref{GEN_ADDITIVITY}.} 
\end{figure}

The relative entropy between two quantum states $x$ and $y$ is defined as
$S(x||y) \equiv - \mbox{tr}(x\log y)-S(x)$,
where $S(x) \equiv - \mbox{tr}(x\log x)$ is the von Neumann entropy of $x$. The relative entropy is a non-negative quantity and due to this property it often appears in the context of distance measure though technically it is not a distance, e.g. it is not symmetric. In Fig. \ref{ALLSTATES}, we present all possible types of correlations present in a quantum state $\rho$. $T_{\rho}$ is the \emph{total mutual information} of $\rho$ given by the distance to the closest product state. If $\rho$ is entangled, its entanglement is measured by the relative entropy of entanglement, $E$, which is the distance to the closest separable state $\sigma$. Having found $\sigma$, one then finds the closest classical state, $\chi_\sigma$, to it. This distance, denoted by $Q$, contains the rest of nonclassical correlations (it is similar to discord \cite{henderson01a, PhysRevLett.88.017901} but entanglement is excluded). We call this quantity \emph{quantum dissonance}. Alternatively, if we are interested in discord \footnote{In this Letter, when we speak of discord we mean the relative entropy of discord as defined by us in Eq. \ref{dis}. When we speak of the original definition discord we will write it as such.}, $D$, then we find the distance between $\rho$ and closest classical state $\chi_{\rho}$. Summing up, we have the following nonclassical correlations:
\begin{align}
E =& \min_{\sigma \in \mathcal{S}} S(\rho || \sigma) 
\quad \textrm{(entanglement)}, \\\label{dis}
D =& \min_{\chi \in \mathcal{C}} S(\rho || \chi) 
\quad \textrm{(quantum discord)}, \\
Q =& \min_{\chi \in \mathcal{C}} S(\sigma || \chi) 
\quad \textrm{(quantum dissonance)}.
\end{align}
Next, we compute classical correlations as the minimal distance between a classically correlated state, $\chi$, and a product state, $\pi$: $C = \min_{\pi\in\mathcal{P}} S(\chi||\pi)$.  Finally, we compute the quantities labeled $L_\rho$ and $L_\sigma$ in Fig. \ref{ALLSTATES}, which give us additivity conditions for correlations.

\emph{Distances.}---We present formulae for the quantities in Fig. \ref{ALLSTATES} beginning with Lemma 1 which describes how we find the closest product state.

\emph{Lemma 1.}
The closest product state of any generic state, $\rho$, as measured by relative entropy, is its reduced states in the product form,  i.e. $\pi_\rho=\pi_1\otimes\dots\otimes\pi_N$.

\emph{Proof.}
Assume that some state, $\alpha=\alpha_1\otimes\dots\otimes\alpha_N$, is the closest product state to $\rho$.  Then consider the difference:
$S(\rho || \pi_\rho)-S(\rho || \alpha)\geq 0$.
Using the linearity of trace and additivity of $\log$ function we have the identity $\tr(\rho\log(\alpha_1\otimes\alpha_2))
=\tr(\tr_2(\rho)\log\alpha_1)+\tr(\tr_1(\rho)\log\alpha_2)$. Applying this identity to both terms of the inequality we have $\sum_i S(\pi_i ||\pi_i)-S(\pi_i || \alpha_i) 0=-S(\pi_\rho||\alpha)\ge 0$, a negative quantity with equality if only if $\pi_\rho =\alpha$. Therefore, for all states $\rho$ we find $\min_{\alpha\in\mathcal{P}} S(\rho || \alpha) = S(\rho ||\pi_\rho).$ $\square$

\emph{Theorem 1.}
The relative entropy of a generic state, $\rho$, and its reduced states in the product form, $\pi_\rho$, is the \emph{total mutual information}.

\emph{Proof.}
Using linearity of trace and additivity of $\log$ we have
$T_\rho\equiv S(\rho||\pi_\rho)= -\tr(\rho\log(\pi_1\otimes\dots\otimes\pi_N) +\rho\log\rho)
=\sum_i-\tr(\pi_i\log\pi_i)+\tr(\rho\log\rho)=S(\pi_\rho)-S(\rho)$, which is \emph{the total mutual information}  (see \cite{tshan} and the references within).  This quantity is equal to the mutual information for bipartite systems. $\square$

\emph{Classical correlations.} The last theorem yields minimal relative entropies for all of the vertical arrows leading to product states in Fig. \ref{ALLSTATES}. Included are also classical correlations given by $C_\sigma=S(\pi_{\chi_\sigma})-S(\chi_\sigma)$ and $C_\rho=S(\pi_{\chi_\rho})-S(\chi_\rho)$.


\emph{Theorem 2.} Given a generic state $\rho$, the closest classical state is $\chi_{\rho}=\sum_{\vec k} \ket{\vec k}\bra{\vec k}\rho \ket{\vec k}\bra{\vec k}$, where $\{\ket{\vec k}\}$ forms the eigenbasis of $\chi_\rho$.

\emph{Proof.} Let $\chi_\rho$ be the closest classical state to $\rho$. Any other classical state $X$ will have more relative entropy (with respect to $\rho$) than $\chi_\rho$; $S(\rho||X)-S(\rho||\chi_\rho)\geq 0$ with equality if and only if $X=\chi_\rho$. Construct $X$ by projecting $\rho$ in the eigenbasis of $\chi_\rho$, $X=\sum_{\vec k} \ket{\vec k}\bra{\vec k}\rho \ket{\vec k}\bra{\vec k}$. Evaluate the first term in the inequality above
by inserting a complete set of projectors $\sum_{\vec k} \ket{\vec k}\bra{\vec k}$ and using the idempotent property of projectors, the cyclic properties of trace, and the fact that $\ket{\vec{k}}\bra{\vec{k}}$ commute with $X$ to obtain
$S(\rho||X)
=-\tr\left(\sum_{\vec k}\ket{\vec k}\bra{\vec k}\rho\log X\right)-S(\rho)
=-\tr\left(\sum_{\vec k}\ket{\vec k}\bra{\vec k}\rho \ket{\vec k}\bra{\vec k} \log X\right)-S(\rho)
=S(X)-S(\rho)$. Using the same techniques the second term of the inequality simplifies as $S(\rho||\chi_\rho)=-\tr(X\log\chi_\rho)-S(\rho)$. Finally the inequality becomes $S(\rho||X)-S(\rho||\chi_\rho) =S(X)-\tr(X\log\chi_\rho) =-S(X||\chi_\rho)\geq 0$, a negative quantity. The only possibility is $-\tr(X\log\chi_\rho)=S(X)=S(\chi_\rho)$, and hence $X=\chi_\rho$. $\square$

\emph{Discord and dissonance.} Theorem 2 yields useful expressions for the quantum discord and quantum dissonance because the minimization of the relative entropy over the classical states is now identical to minimization of the entropy $S(\chi_x)$ over the choice of local basis $\ket{\vec k}$:
\begin{gather}\label{discordformual}
D=S(\chi_\rho)-S(\rho) \quad\mbox{and}\quad Q=S(\chi_\sigma)-S(\sigma),
\end{gather}
where $S(\chi_x)=\min_{\ket{\vec k}}S\left(\sum_{\vec k}\ket{\vec k}\bra{\vec k}x\ket{\vec k}\bra{\vec k}\right).$


\emph{Theorem 3.}
The equations for $L_\rho$ and $L_\sigma$ are
$L_\rho= S(\pi_{\chi_\rho})-S(\pi_\rho)$ and $L_\sigma= S(\pi_{\chi_\sigma})-S(\pi_\sigma).$

\emph{Proof.}
To find the formula for $L_\rho$ we start by evaluating $S(\rho||\pi_{\chi_\rho})=-\tr(\rho\log\pi_{\chi_\rho})-S(\rho)$.
Using the fact that $\pi_{\chi_\rho}$ has the same basis as $\chi_\rho$ and inserting a complete set of projectors in that basis in the first term gives us
$-\tr(\rho\log\pi_{\chi_\rho})=-\tr(\chi_\rho\log\pi_{\chi_\rho})$.  The additivity of the log and linearity of  trace gives $-\tr(\rho\log\pi_{\chi_\rho})=S(\pi_{\chi_\rho})$. 
On the other hand we can use the linearity of trace right away to get
$-\tr(\rho\log\pi_{\chi_\rho})=-\tr(\pi_\rho\log\pi_{\chi_\rho})=S(\pi_{\chi_\rho})$ or $\pi_{\chi_\rho}=\ket{\vec k}\bra{\vec k}\pi_\rho\ket{\vec k}\bra{\vec k}$, where $\{\ket{\vec k}\}$ forms the basis of $\chi_\rho$. Finally
$L_\rho=S(\pi_\rho||\pi_{\chi_\rho})=S(\pi_{\chi_\rho})-S(\pi_\rho)$. The proof for the `$\sigma$' side proceeds in the same way. $\square$

The theorems above give us a method to compute all classical and quantum correlations other than entanglement.  Surprisingly, they also give us the following additivity relations for correlations:
\begin{gather}
T_\rho  = D + C_\rho - L_\rho, \quad \mbox{and} \quad T_\sigma = Q + C_\sigma - L_\sigma.
\label{GEN_ADDITIVITY}
\end{gather}
These relations correspond to the closed paths in Fig. \ref{ALLSTATES} and mean that the sum of quantum and classical correlations is equal to the sum of total mutual information and the quantities labeled as $L_\rho$ and $L_\sigma$. Though, there is no physical interpretation for $L_\rho$ and $L_\sigma$, yet these quantities play a role in forming relations such as above.  Note, above entanglement is present `within' discord but not  by itself. We may wonder how do entanglement, dissonance, and classical correlations compare to the total mutual information.  

\emph{Examples.}---We offer three examples (two with multipartite states) below in which we calculate all possible correlations and find the additivity relations.


\emph{1. Bell-diagonal states.--}
Consider mixed states of two qubits with vanishing Bloch vectors for the reduced operators. They are equivalent up to local unitary operations to Bell diagonal states
$\rho = \sum_{i = 1}^4 \lambda_i \ket{\Psi_i} \bra{\Psi_i}$, where $\lambda_i$ are ordered in non-increasing size
and $\ket{\Psi_i}$ are the four Bell states. $\rho$ is entangled when $\lambda_1>1/2$. The closest separable state is $\sigma = \sum_{i=1}^4 p_i \ket{\Psi_i} \bra{\Psi_i}$ where $p_1=1/2$ and the remaining probabilities are $p_i=\lambda_i/(2(1-\lambda_1))$ \cite{VP}. Following the calculation for $\sigma$ in \cite{VP} one can show the closest classical states are given by $\chi=\tfrac{q}{2}\left[\ket{\Psi_1}\bra{\Psi_1}+\ket{\Psi_2}\bra{\Psi_2}\right] + \tfrac{1-q}{2}\left[\ket{\Psi_3}\bra{\Psi_3}+\ket{\Psi_4}\bra{\Psi_4}\right]$,
with $q_{\rho}=\lambda_1+\lambda_2$ and $q_{\sigma}=p_1+p_2$. The product states $\pi$ are all identical and given by the normalized identity $\openone/4$. Given these states one can calculate entanglement, discord, dissonance and classical correlations. The correlations are subadditive: $T_\rho \geq E + Q+ C_\sigma$.

\emph{2. $W$ state.--} The closest separable state to a bipartite pure entangled state is a classical state \cite{VP}.  This means entanglement represents all quantum correlations leading to the additivity relation $T_\rho=E+C_\rho$. Multipartite \emph{pure} states may contain other nonclassical correlations than entanglement. Consider $\ket{W}$ state of three qubits $\ket{W} = \tfrac{1}{\sqrt{3}} \left( \ket{100} + \ket{010} + \ket{001} \right)$. The state is clearly entangled with the closest separable state of the form \cite{WEGM}, $\sigma = \tfrac{8}{27} \ket{000}\bra{000} + \tfrac{12}{27} \ket{W} \bra{W} + \tfrac{6}{27} \ket{\overline{W}} \bra{\overline{W}} + \tfrac{1}{27} \ket{111} \bra{111}$, where $\ket{\overline{W}} = \tfrac{1}{\sqrt{3}}(\ket{011} + \ket{101} + \ket{110})$. Contrary to the bipartite case, $\sigma$ is not a classically correlated state. Moreover $\chi_\rho$ and $\chi_\sigma$ are different states obtained by dephasing $\rho$ in the standard basis for $\chi_\rho$ and $\sigma$ in the $x$ basis for $\chi_\sigma$. The correlations in the $\ket{W}$ are: $E \approx 1.17$, $D \approx 1.58$,    $Q \approx 0.94$, $C_\rho\approx 1.17$, $C_\sigma\approx 0.36$, $L_\rho=0$, and $L_\sigma=0.24$.  Once again this gives us subadditivity of correlations: $T_\rho>E+Q+C_\sigma$.  Entanglement and dissonance are said to jointly belong to discord, but when combined the two are greater than discord in this example, $D<E+Q$.

\emph{3. Cluster state.--} Cluster state is a pure multipartite state which has been known as a useful resource for measurement based quantum computation \cite{Briegel01, clustercomp}. Cluster state for four parties is $\ket{C_{4}} =\ket{0+0+} +\ket{1+1+} +\ket{0-1-} +\ket{1-0-}$. The closest separable state to the cluster state is $\sigma_{C_{4}} = \frac{1}{4} \big( \ket{0+0+}\bra{0+0+} + \ket{1+1+}\bra{1+1+} + \ket{0-1-}\bra{0-1-} + \ket{1-0-}\bra{1-0-} \big)$ \cite{WEGM}, which is a classical state. The correlations are: entanglement is equal to discord, $E=2$, $Q=0$, and $C_\rho=2$ and we also have the additivity relation $T_\rho=E+C_\rho$. It is surprising, in light of the previous example, that the correlations in a cluster state behave like the correlations in pure bipartite states.

From the examples above we conjecture that the correlations of a quantum state are subadditive in the sense $T_\rho\geq E+Q+C_\sigma$.  The source of the subadditivity may be due to entanglement being less than the difference in the entropies of $\rho$ and $\sigma$, i.e. $S(\sigma)\geq-\tr(\rho\log\sigma)$.  We have not been able to prove this explicitly nor have we found an example showing the contrary.

\emph{Comparison with other measures.}---We now compare our measure of quantum correlations with two other measures of nonclassicality, the original quantum discord and measurement induced disturbance.

The original definition of discord \cite{henderson01a,PhysRevLett.88.017901} involves bipartite systems with classicality for only one subsystem.  We can define a classical state in this manner by restricting the projective operation to one subspace as $\chi_\rho=\sum_{k} \ket{k}\bra{k}\otimes\openone \rho \ket{k} \bra{k}\otimes\openone$.  This does not alter any theorems of this paper. Avoiding minimization for the moment, the equation for original discord is $\mathcal{\delta} =S(\rho_A) -S(\rho) + \sum_k p_{k} S(\ket{k}\bra{k}\otimes\rho_B^{k})$. Using $S\left(\sum_k p_k \ket{k}\bra{k} \otimes\rho_B^k\right) =S(\sum_k p_k \ket{k}\bra{k})+\sum_k p_k S(\ket{k}\bra{k}\otimes\rho_B^k)$ with
$S(\sum_k p_k\ket{k}\bra{k})=S(\tr_B(\chi_\rho))$ and adding and subtracting $S(\rho_B)=S(\tr_A(\chi_\rho))$ we get $\delta= S(\pi_\rho) -S(\rho) +\left(S(\chi_\rho) -S(\pi_{\chi_\rho})\right)=D-L_\rho$ or equivalently $\delta=T_\rho-C_\rho$. This is a remarkably simple relationship between the two forms of discord with the key difference being in minimization. We minimize the quantity $D$, while for the original discord, $D-L_\rho$ is minimized over all measurements $\ket{k}\bra{k}$. Also note that this relation may not hold when the original discord is considered with positive operator values measure as the analysis above only considers projective operations. 

\emph{Measurement induced disturbance} (MID) \cite{luo:022301} is defined as the difference in the mutual information of $\rho$ and $\eta=\sum_{ij} \ket{ij} \bra{ij} \rho\ket{ij}\bra{ij}$, where $\{\ket{ij}\}$ form the basis of the product state $\pi_\rho$.  This means both $\rho$ and $\eta$ have the same reduced states, leading to the formula $I(\rho)-I(\eta)=S(\eta)-S(\rho)$ for MID. Once again this is remarkably similar to discord (or dissonance) defined in the Letter.  The difference between the two measures is the minimization, as the basis of $\pi_\rho$ may not minimize the relative entropy, i.e. $S(\eta)\geq S(\chi_\rho)$.  This can be seen in the following nonclassical two qubit state: $\rho=(1-q)\sum_{ij}p_{ij}\ket{ij}_{zz} \bra{ij} +q \sum_i \frac{1}{2} \ket{ii}_{xx} \bra{ii}$.  The reduced states of $\rho$ are diagonal in the $z$ basis and therefore $\eta$ is given by the diagonal elements of $\rho$ in that basis.  For the values of $p_{ij}$ near $\frac{1}{4}$, $\eta$ will nearly be a fully mixed state, but if the value of $q$ is large enough then the entropy of $\chi_\rho$ will be minimized in a basis that is close to the $x$ basis.  This shows that MID is not the same as relative entropy of discord. When $L_\rho=0$, the reduced basis of $\rho$ and $\chi_\rho$ are the same, and therefore for those states MID is the same as our discord.

\emph{Conclusions.}---We have discussed the problem of separation of total correlations in a given quantum state into quantum entanglement, dissonance, and classical correlations. Quantum entanglement and dissonance, whose novel definition is introduced here, jointly belong to what is known as quantum discord. Putting all correlations on an equal footing has another potential advantage in addition to those discussed in our work. Namely, given that relative entropy between two states tells us how distinguishable they are \cite{PhysRevA.56.4452}, a question arises as to whether this quantity is connected to the efficiency of quantum information processing. Can dissonance, for instance, give us more efficient information processing to what classical correlations allow? Even more fascinatingly, could it be that dissonance is as powerful as entanglement as far as  quantum computing is concerned? We hope that our definitions of correlations, which apply to any number of subsystems of arbitrary dimensionality, will provide further stimulus for these important and fundamental questions.

\begin{acknowledgements}
We acknowledge the financial support by the National Research Foundation and the Ministry of Education of Singapore. K.M. thanks L. Aolita and C. Rodr\'iduez-Rosario for helpful conversations.
\end{acknowledgements}

\bibliography{gencorr.bib}

\end{document}